\documentclass[twocolumn,10pt,showpacs,pra,floatfix,superscriptaddress]{revtex4-1}
\usepackage{bm}
\usepackage{amssymb}
\usepackage{amsmath}
\usepackage{graphicx}
\usepackage{color,soul}
\usepackage{threeparttable}

\allowdisplaybreaks

\begin{document}
\title{Exploiting transport properties for the detection of optical pumping in heavy ions}
\author{Mustapha~Laatiaoui }
\email{mlaatiao@uni-mainz.de}
\affiliation{Department Chemie, Johannes
Gutenberg-Universit\"at, Fritz-Strassmann Weg 2, 55128 Mainz,
Germany} \affiliation{Helmholtz-Institut Mainz, Staudingerweg 18,
55128 Mainz, Germany} \affiliation{GSI
Helmholtzzentrum f\"ur Schwerionenforschung GmbH, Planckstrasse 1,
D-64291 Darmstadt, Germany} \affiliation{KU Leuven, Instituut voor
Kern- en Stralingsfysica, Celestijnenlaan 200D, B-3001 Leuven,
Belgium}
\author{Alexei~A.~Buchachenko}
\affiliation{CEST, Skolkovo Institute of Science and Technology, Skolkovo
Innovation Center, Nobel str. 3, Moscow 121205, Russia}
\affiliation{Institute of Problems of Chemical Physics RAS,
Chernogolovka, Moscow District 142432, Russia}
\author{Larry~A.~Viehland }
\affiliation{Science Department, Chatham University, Pittsburgh,
Pennsylvania 15232, USA}
\date{\today}

\begin{abstract}
We present a kinetic model for optical pumping in Lu$^+$ and Lr$^+$
ions as well as a theoretical approach to calculate the transport
properties of Lu$^+$ in its ground $^1S_0$ and metastable
$^3D_1$ states in helium background gas. Calculations of the
initial ion state populations, the field and temperature dependence
of the mobilities and diffusion coefficients, and the ion arrival
time distributions demonstrate that the ground- and metastable-state
ions can be collected and discriminated efficiently under realistic
macroscopic conditions.
\end{abstract}

\maketitle
\section{Introduction}
Atomic structure studies complement element discoveries and advance
our understanding of atoms and their
nuclei~\cite{Blaum:2013,Campbell:2016}. In
Ref.~\cite{Laatiaoui:2020}, a new method of optical spectroscopy is
being developed for such studies on element cations beyond nobelium
($Z>102$), which are currently inaccessible by common techniques.
Optical pumping utilizing laser resonant excitations is envisaged to
populate metastable electronic states of ions, which can be then
discriminated from ions in the ground state utilizing electronic
state chromatography~\cite{Kemper:1991,Taylor:1999}. The latter
technique exploits the fact that monoatomic ions in different
electronic states experience different interactions in ion-atom
collisions. In particular, a change in the electronic configuration,
such as that encountered when metastable states are occupied, results in
different transport coefficients (ion mobility and diffusion
coefficients), which control the rate of the field-induced ion drift
through an inert gas~\cite{Laatiaoui:2012}. While the difference in
ion-atom interaction reflects the fundamental electronic structures,
the drift time of ions is controlled by macroscopic parameters
(temperature, pressure, and electric-field strength), which can be
optimized to achieve the best collection or discrimination of the
ions. The use of electronic state
chromatography in conjunction with ablation sources for
state-selected ion chemistry is an established
technique~\cite{Armentrout:2011}. Such studies
can span a variety of elemental cations along the first-, second-, and
third-row transition metals for which state-specific mobilities have
already been measured~\cite{Manard:2016,Manard:2016b,Iceman:2007}.
Importantly, in this context, electronic structure information can be
extracted while searching for suitable ground-state transitions for
resonant optical pumping (see Sec.~\ref{Sec_Pumping}).

In the present work, we provide the theoretical
proof of the optical spectroscopy concept, called laser resonance
chromatography~\cite{Laatiaoui:2020}, for Lu$^+$ and its heavier chemical homologue,
Lr$^+$ ($Z=103$). Following the concept, a few ions are first bunched inside a radiofrequency buncher for optical pumping and then released to a drift tube for electronic-state chromatography. In the following Sec.~\ref{Sec_Pumping}, we propose a simple resonant pumping scheme allowing for an efficient population of the metastable states of the
ions. We predict the interaction potentials of Lu$^+$ in the
metastable state with helium in Sec.~\ref{Sec_Pots} and describe the gaseous ion transport in ground and metastable states at different conditions of temperature $T$
and ratio of electric-field strength to gas number density $E/n_0$ in Sec.~\ref{Sec_Transport}.
Based on the analysis of expected arrival time distributions introduced in Sec.~\ref{Sec_Simulations}, we identify parameter ranges for electronic state chromatography in Sec.~\ref{Sec_Signal} and
deduce achievable ion transmission and collection efficiencies, and
the efficiency for detecting resonant optical excitations.

\section{Optical pumping in singly-charged lutetium and lawrencium}
\label{Sec_Pumping} 
\begin{figure}
\includegraphics[scale=0.33]{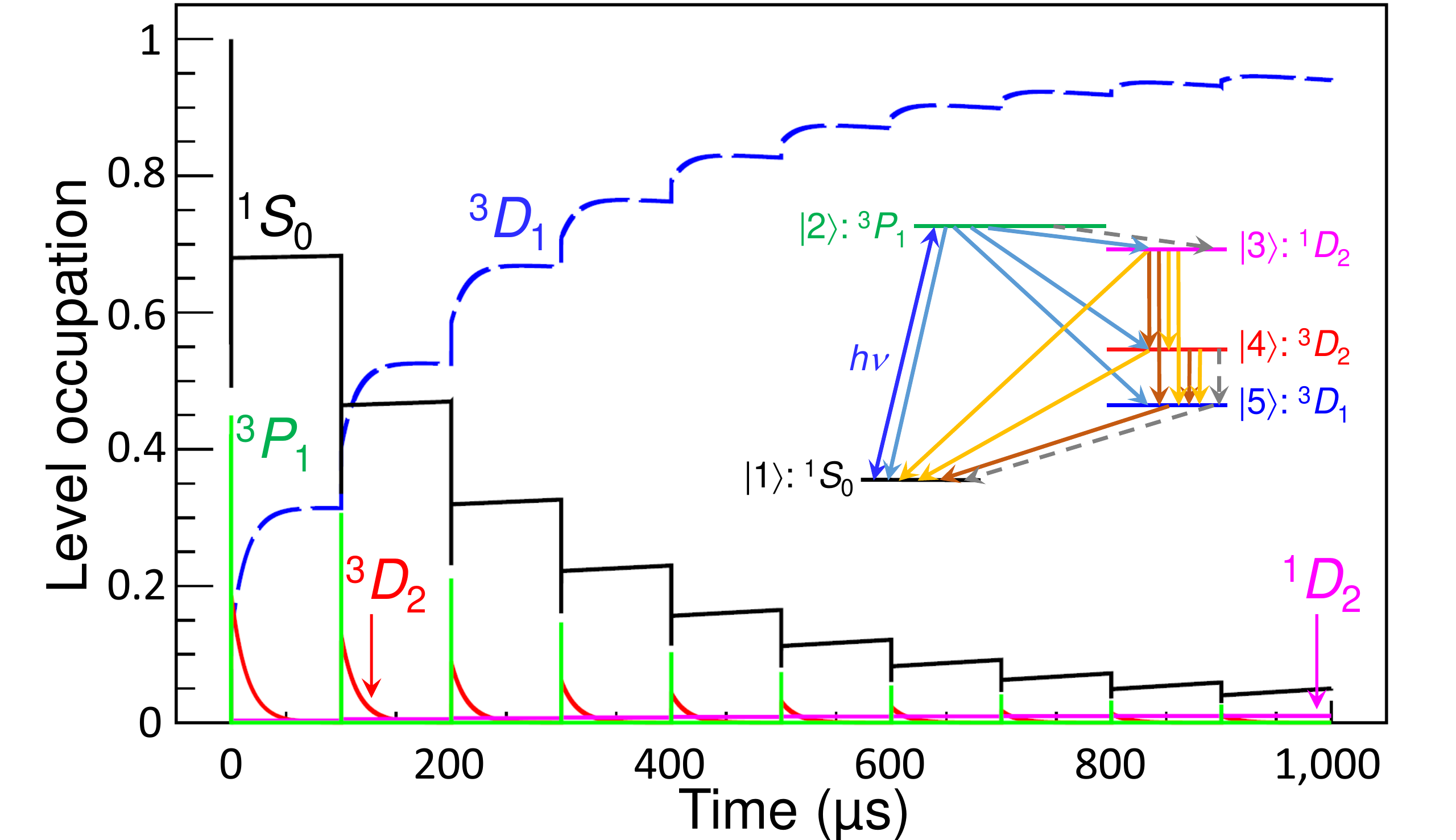}
\caption{\label{fig:Pump} Laser-induced population
transfer from ground- to metastable states in Lu$^+$ at
$5\times10^{-2}\,$mbar He. Level occupation is indicated for each of
the modeled states $|i\rangle$ in the course of $10$ laser beam
exposures. Inset: corresponding five-level system used in the
rate-equation model with arrows in blue, yellow, and brown
indicating $E1$, $E2$, and $M1$ transitions, respectively. Laser probing
($h\nu$) of the intermediate $^3P_1$ state induces optical pumping
in the system. Collision-induced relaxation is marked by dashed arrows.}
\end{figure}

We developed a rate equation model for a
five-level system to evaluate optical pumping in Lu$^+$ (Lr$^+$) prior to
electronic state chromatography. The system consists of the ground
state $|1\rangle$: $6s^2\,^1S_0$ ($7s^2\,^1S_0$), the
intermediate level $|2\rangle$: $6s6p\,^3P_1$ ($7s7p\,^3P_1$)
that should be probed by laser radiation, and three low-lying
metastable states $|3\rangle$: $6s5d\,^1D_2$ ($7s6d\,^1D_2$),
$|4\rangle$: $6s5d\,^3D_2$ ($7s6d\,^3D_2$), and $|5\rangle$:
$6s5d\,^3D_1$ ($7s6d\,^3D_1$)~\cite{NIST:2018,Kahl:2019}, which
serve to collect the population from $|2\rangle$ by radiative and
collision-induced relaxation processes~\cite{Laatiaoui:2020} (see
inset in Fig.~\ref{fig:Pump}). Collisional quenching was considered
for $^3P_1$ to $^1D_2$, $^3D_2$ to $^3D_1$, and $^3D_1$ to
$^1S_0$ at different gas number densities $n_0$ by including the
corresponding quenching rates as reported for the isoelectronic
neutral barium in He at $880\,$K,
$\alpha_{23}/n_0=8\times10^{-11}\,$cm$^3$/s,
$\alpha_{45}/n_0=6\times10^{-11}\,$cm$^3$/s, and
$\alpha_{51}/n_0=10^{-13}\,$cm$^3$/s,
respectively~\cite{Brust:1995}. In addition, it is assumed that
broadband laser radiation is used during the initial level search
such that the coherence terms in the optical Bloch equations can be
safely neglected~\cite{Loudon:2000}. We obtain
\begin{eqnarray}
  \nonumber\frac{d\rho_1}{dt} &=& A_{21}\rho_2 + A_{31}\rho_3 + A_{41}\rho_4 + (A_{51} +
  \alpha_{51})\rho_5 \\
                              &-& \frac{1}{2}A_{21}S(\omega_L,\omega_{12})O(t)(\rho_1 - \rho_2) \\
  \nonumber\frac{d\rho_2}{dt} &=& \frac{1}{2}A_{21}S(\omega_L,\omega_{12})O(t)(\rho_1 - \rho_2) \\
                              &-& (A_{21} + A_{23} + A_{24} + A_{25} + \alpha_{23})\rho_2 \\
  \nonumber\frac{d\rho_3}{dt} &=& (A_{23} + \alpha_{23})\rho_2 \\
                              &-& (A_{31} + A^e_{34} + A^m_{34} + A^e_{35} + A^m_{35})\rho_3 \\
  \nonumber\frac{d\rho_4}{dt} &=& A_{24}\rho_2 + (A^e_{34} + A^m_{34})\rho_3 \\
                              &-& (A_{41} + A^e_{45} + A^m_{45} + \alpha_{45})\rho_4 \\
  \nonumber\frac{d\rho_5}{dt} &=& A_{25}\rho_2 + (A^e_{35} + A^m_{35})\rho_3 \\
                              &+& (A^e_{45} + A^m_{45} + \alpha_{45})\rho_4 - (A_{51} +
                              \alpha_{51})\rho_5
\end{eqnarray}
with the normalization $\sum_i \rho_i = 1$ and the initial
conditions $\rho_1(t=0)=1$ and $\rho_i(t=0)=0$ for $1<i$, where
$\rho_i$ with $i=1...5$, correspond to the occupations of individual states
$|i\rangle$. $A_{ki}$, $A^e_{ki}$, and $A^m_{ki}$ are the Einstein
coefficients for spontaneous emission from $|k\rangle$ to
$|i\rangle$ via $E1$, $E2$, and $M1$ transitions,
respectively~\cite{Quinet:1999,NIST:2018,Paez:2016,Kahl:2019}.

We used the frequency-dependent saturation parameter
$S(\omega_L,\omega_{12})$ as described in Ref.~\cite{Chhetri:2017}.
In this parameter, we considered Doppler broadening at room
temperature in terms of full width at half maximum of $0.8\,$GHz
($0.7\,$GHz) for Lu$^+$ (Lr$^+$) and the spectral bandwidth
of the laser of $2.5\,$GHz, as well as dephasing effects on the order of $1.4\,$GHz from mode
fluctuations within the laser pulse~\cite{Chhetri:2017}. For both ionic species, an energy
density of the laser radiation of $10\,\mu$J/cm$^2$ was taken.
For simplicity's sake, we neglected hyperfine structures and nuclear
isomerism as these should be covered by the broadband laser radiation.
In addition, we neglected broadening effects from radial macro-motion of ions
as well as collisional dephasing effects because they do not affect the
results at the expected background pressures.
Moreover, we included a rectangular function $O(t)$ into the model
to mimic laser pulse exposures of $10\,$ns
duration and $100\,\mu$s period.

Figure~\ref{fig:Pump} shows the calculated level
occupations in Lu$^+$ in the course of $10$ laser
exposures at a He background pressure of
$5\times10^{-2}\,$mbar. The efficiency for optical pumping from the
ground state into the $^3D_1$ level is $31.5$\% for one laser
pulse exposure and reaches a
value of $94$\% for $10$ laser pulses. Similar calculations for
Lr$^+$ indicate a lower but still sufficiently high efficiency of
$7.5$\% and $53$\% for one and $10$ pulse
exposures, respectively.

In addition, a substantial population transfer in both ionic species
can be obtained to the $^1D_2$ ($^3D_2$) level at
$5\times10^{-1}\,$mbar ($5\times10^{-3}\,$mbar) background
pressures. Collisional quenching is found to depopulate the
$^3D_1$ state only slowly at $5\times10^{-1}\,$mbar and is a
negligibly small effect for pressures $\leq 5\times10^{-2}\,$mbar.

\section{$\rm{Lu}^+$--$\rm{He}$ interaction potentials}
\label{Sec_Pots} The scalar-relativistic (SR) interaction potentials
between the ground-state Lu$^+$($6s^2\,^1S$) ion and rare gas atoms
from He to Xe were calculated {\it ab initio} in
Ref.~\cite{Buchachenko:2014}. The reduced zero-field mobility in He
at room temperature computed as described in Sec.~\ref{Sec_Transport}
below is $K_0=16.578\,$cm$^2$/Vs. This was later confirmed
experimentally as $K_0= 16.8\pm0.4\,$cm$^2$/Vs \cite{Manard:2017}.
Good agreement with the experimental data was also found for other
lanthanide ions studied: Yb$^+$, Eu$^+$ and Gd$^+$
\cite{Buchachenko:2014,Manard:2017,Laatiaoui:2012,Buchachenko:2019}.
We therefore used similar theoretical techniques to address the
interaction of the metastable Lu$^+$($6s5d\,^3D$) ion with He.

In brief, the small-core effective core potential (ECP) ECP28MWB
\cite{DOLG:1989} and  the segmented basis set \cite{Cao:2002}
augmented by the $s2pdfg$ set of primitive diffuse functions
\cite{Buchachenko:2007} were employed for Lu, whereas the
aug-cc-pV5Z basis was used for He \cite{Woon:1994}. The $3s3p2d2f1g$
bond function set~\cite{Cybulski:1999} was placed in the middle of
the Lu--He distance. The ground-state $X^1\Sigma^+$ SR potential was
computed as in Ref.~\cite{Buchachenko:2014} using the restricted
coupled-cluster method with singles, doubles, and noniterative
triples, CCSD(T)~\cite{KNOWLES:1993, KNOWLES:2000}, as implemented
in the MOLPRO program package~\cite{MOLPRO:2015} for the restricted
Hartree-Fock reference. For the metastable state, the restricted
Hartree-Fock wave functions were calculated by fixing the single
occupation of the particular Lu$^+$ $5d_0\sigma$ or $5d_{+2}\delta$
molecular orbital, allowing us to resolve $^3\Sigma^+$ and
$^3\Delta$ molecular states that fall in the same $A_1$
representation of the $C_{2v}$ symmetry group. The CCSD(T) method
was then implemented for each reference wave function, with the
Lu$^+$ $4s^24p^64d^{10}$ shells kept as core and the counterpoise
correction \cite{Boys:1970} applied individually to each state. This
state-resolved approach was proven to be successful for the
Gd$^+$($^{10}D^\circ$)--He, Ar
interactions~\cite{Buchachenko:2019}. Table~\ref{tab:pots} presents
the equilibrium parameters, distance $R_e$ and well depth $D_e$, and
dissociation energy $D_0$ obtained in the SR CCSD(T) calculations.
\begin{center}
\begin{table}
%\centering
\caption{Equilibrium distances $R_e$, well depths $D_e$, and dissociation energies $D_0$ of the
Lu$^+$--He potentials corresponding to the ground $^1S$ ($^1S_0$)
and metastable $^3D$ ($^3D_1$) states of the ion.} \label{tab:pots}
\begin{threeparttable}
\begin{tabular}{l@{\qquad}c@{\qquad}c@{\qquad}c@{\qquad}}
\hline \hline
State & $R_e$ ({\AA}) & $D_e$ (cm$^{-1}$) & $D_0$ (cm$^{-1}$) \\
\hline
\multicolumn{4}{c}{Scalar relativistic} \\
%\hline
$X^1\Sigma^+$($^1S$)  & 4.17 & 47.3 & 31.4 \\
$^3\Sigma^+$($^3D$) & 4.41 & 32.6 & 20.3 \\
$^3\Pi$($^3D$) & 3.77  & 61.0 & 43.9 \\
$^3\Delta$($^3D$) & 3.83 & 62.8 & 45.2 \\
%\hline
\multicolumn{4}{c}{SO coupled} \\
%\hline
$X0^+$($^1S_0$) & 4.17 & 47.3 & 31.4 \\
$0^-_1$($^3D_1$) & 4.11 & 49.9 & 34.2 \\
$1_1$($^3D_1$) & 3.91 & 52.2 & 36.4 \\
\hline \hline
\end{tabular}
\end{threeparttable}
\end{table}
\end{center}
To take into account the spin-orbit (SO) coupling that determines
the fine structure of the metastable state, the state-interacting SO
configuration-interaction method~\cite{BERNING:2000} was employed.
The ECP description gives poor results for SO
coupling~\cite{Buchachenko:2019}, so we resorted to the all-electron
description. It was found that the measured fine-structure
splittings~\cite{NIST:2018} cannot be reproduced well for isolated
$^3D$ multiplet. Test calculations indicated that improvement can be
achieved by taking into account the higher lying $6s5d\,^1D$ term.
The SR potentials correlating to this term were computed using the
multireference configuration-interaction
method~\cite{WERNER:1988,SHAMASUNDAR:2011} with the state-averaged
complete active space multiconfigurational self-consistent field
reference orbitals~\cite{WERNER:1985} and the ECP approach described
above. The resulting SR potentials together with the CCSD(T)
potentials for the ground and metastable triplet states were taken
as the diagonal part of the SO Hamiltonian matrix. The Breit-Pauli
coupling matrix elements were calculated adopting the all-electron
X2C (``exact two-component'')~\cite{PENG:2012} SR approximation with the
cc-pwCVDZ and aug-cc-pVDZ basis sets for Lu and He,
respectively~\cite{Lu:2016,Woon:1994}. In all multireference
calculations, Lu$^+$ $6s5d$ orbitals were considered as active,
while the Lu$^+$ $5p4f$ and He $1s$ orbitals were correlated as
doubly occupied.

\begin{figure}
\includegraphics[scale=0.3]{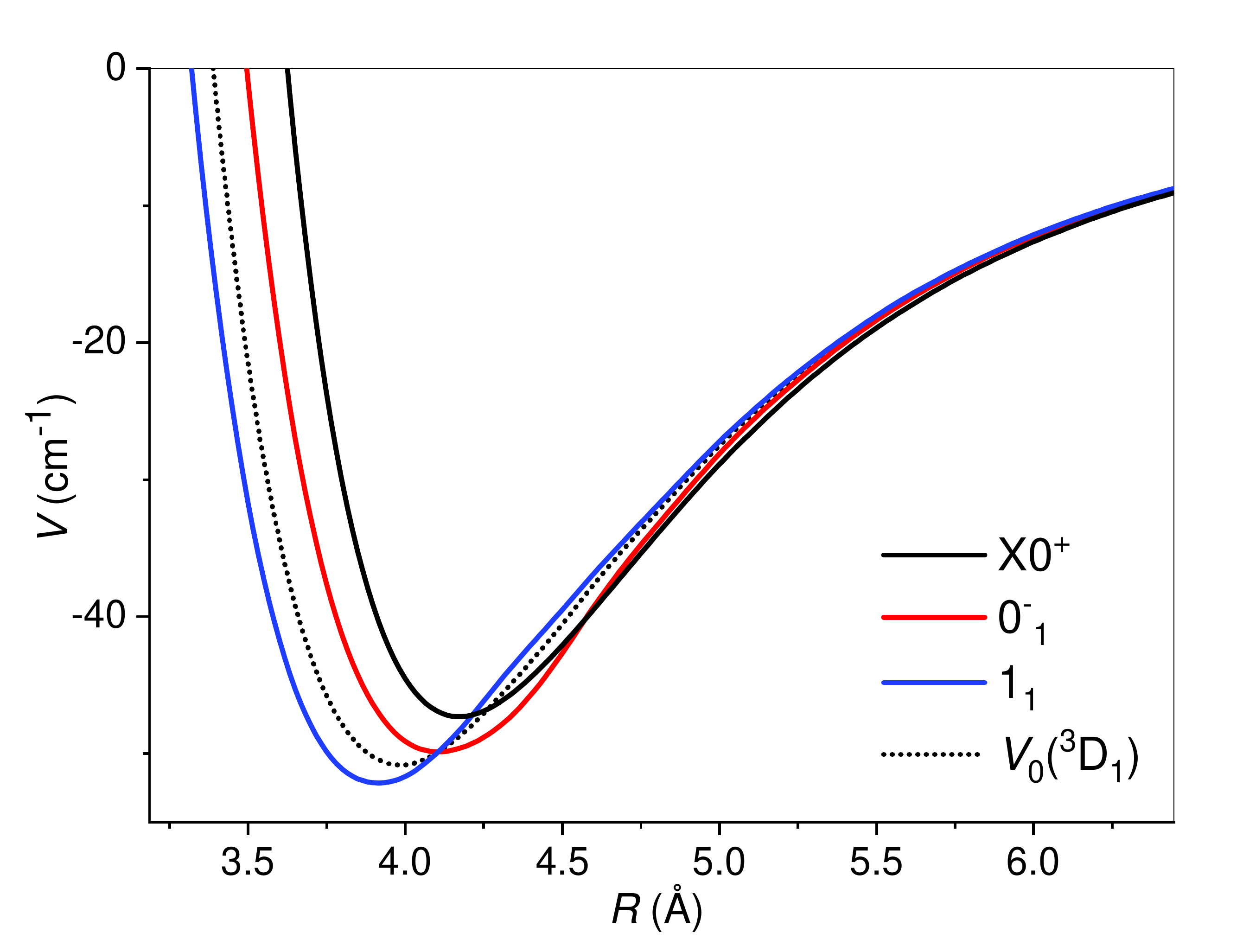}
\caption{\label{fig:SOpot} Lu$^+$--He SO-coupled interaction
potentials corresponding to the $^1S_0$ and $^3D_1$ states of the
ion. Dotted line represents isotropic interaction potential for the
$^3D_1$ state. For the sake of comparison, all potentials are
referred to the same dissociation limit.}
\end{figure}

The resulting energies of the $^3D_J$ fine-structure levels $J=2$,
$3$ with respect to the one with $J=1$ are $639$ and $2236\,$cm$^{-1}$,
which are in good agreement with the experimental values of $639$
and $2403\,$cm$^{-1}$~\cite{NIST:2018}, respectively. The
SO-coupled potentials relevant to the transport calculations are
shown in Fig.~\ref{fig:SOpot}, while their parameters are presented
in Table~\ref{tab:pots}. In the $\Omega^\sigma_J$ notations of the
Hund case (c) coupling scheme, with $J$ and $\Omega$ being the total
angular electronic momentum and its projection onto the interatomic
axis, respectively, and $\sigma$ being inversion parity, these
states are $X0^+$, which replaces the ground $X^1\Sigma^+$ SR state,
and 0$^-_1$, 1$_1$, which correlate to the $^3D_1$ Lu$^+$ term.
One should note that the He interaction with the metastable Lu$^+$
ion is stronger than that with the ground-state ion.

\section{Transport coefficients}
\label{Sec_Transport} Transport properties were calculated for the
$^{175}$Lu$^+$ ion using the Gram-Charlier approach to solving the
Boltzmann equation~\cite{VIEHLAND:1994,Viehland:2018,LXCAT:2019}. For the
ground-state ion, the momentum-transfer and other transport cross
sections were calculated as functions of the collision energy for the
single $X0^+$ interaction potential. For the metastable
Lu$^+$($^3D_1$) ion, we used so-called ``anisotropic SO-coupled
approximation''~\cite{Buchachenko:2014,Buchachenko:2019,LXCAT:2019},
in which the cross sections for the $0^-_1$ and $1_1$ interaction
potentials were averaged with the degeneracy factors $1/3$ and
$2/3$, respectively.
\begin{figure}
\includegraphics[scale=0.33]{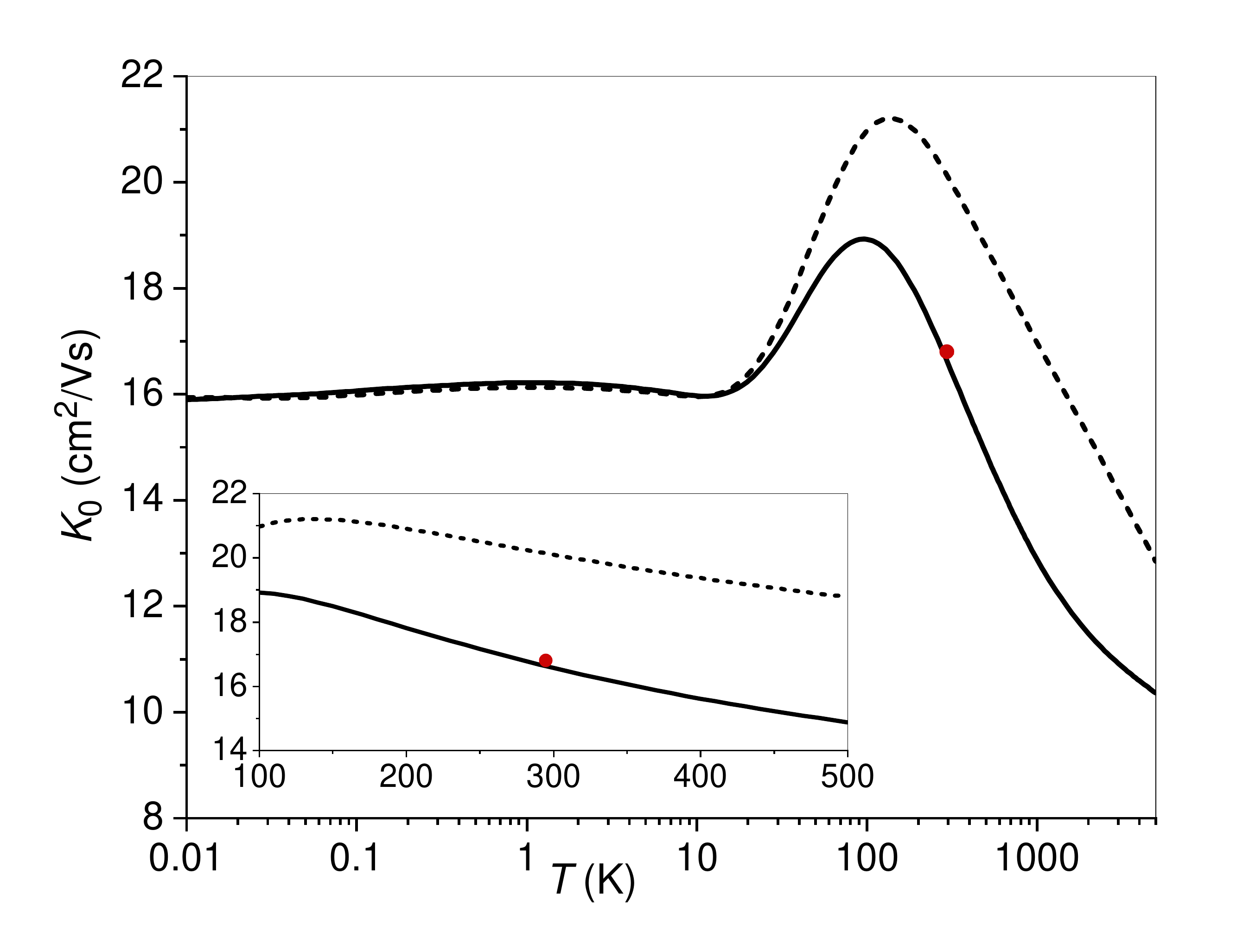}
\caption{\label{fig:K0vsT} Reduced zero-field mobilities of the
Lu$^+$ ions in the ground $X0^+$ state (solid line) and metastable
$^3D_1$ state (dashed line) as functions of temperature. The dot
indicates the experimental result for the ground
state~\cite{Manard:2017}. The inset provides an enlarged view of the
$100-500\,$K region.}
\end{figure}
\begin{figure}
\includegraphics[scale=0.33]{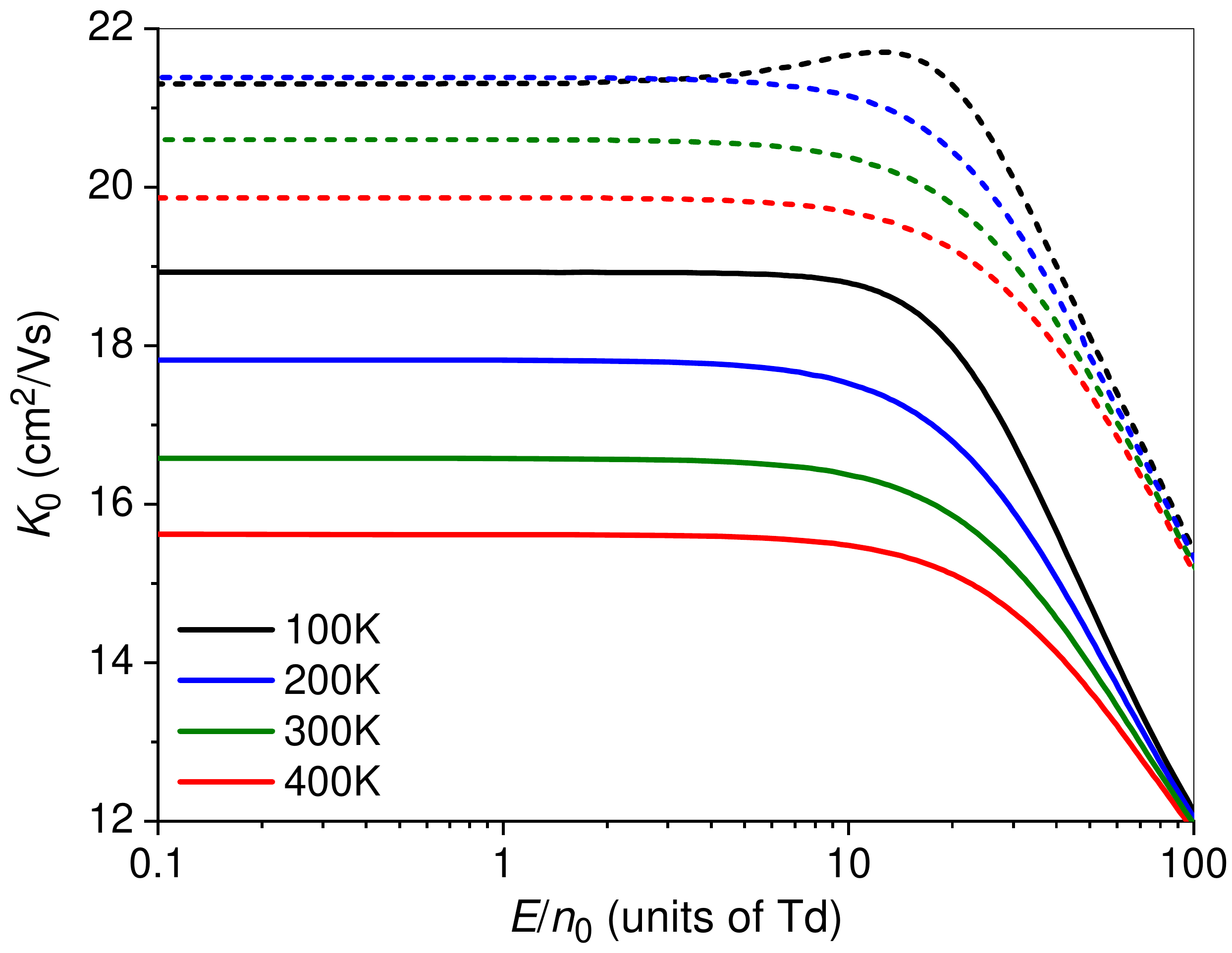}
\caption{\label{fig:K0vsn0} Reduced mobilities of the Lu$^+$ ions in
the ground $X0^+$ state (solid lines) and metastable $^3D_1$ state
(dashed lines) as functions of $E/n_0$ at selected temperatures. }
\end{figure}

Figure~\ref{fig:K0vsT} shows the temperature dependence of the
reduced zero-field mobilities for the ions in two electronic states,
while the reduced mobility dependence on the ratio of the field
strength to gas number density is presented in
Fig.~\ref{fig:K0vsn0}. Effects of temperature and $E/n_0$ are quite
remarkable. With increasing $T$, deviations in zero-field mobilities
of up to $24$\% can be reached. A so-called low-field region with
constant mobility holds below $E/n_0\approx10\,$Td, with $1\,$Td
(Townsend unit) being equal to $10^{-21}\,$Vm$^2$; then the mobilities
of ions in both states start to decline.

\begin{figure}
\includegraphics[scale=0.33]{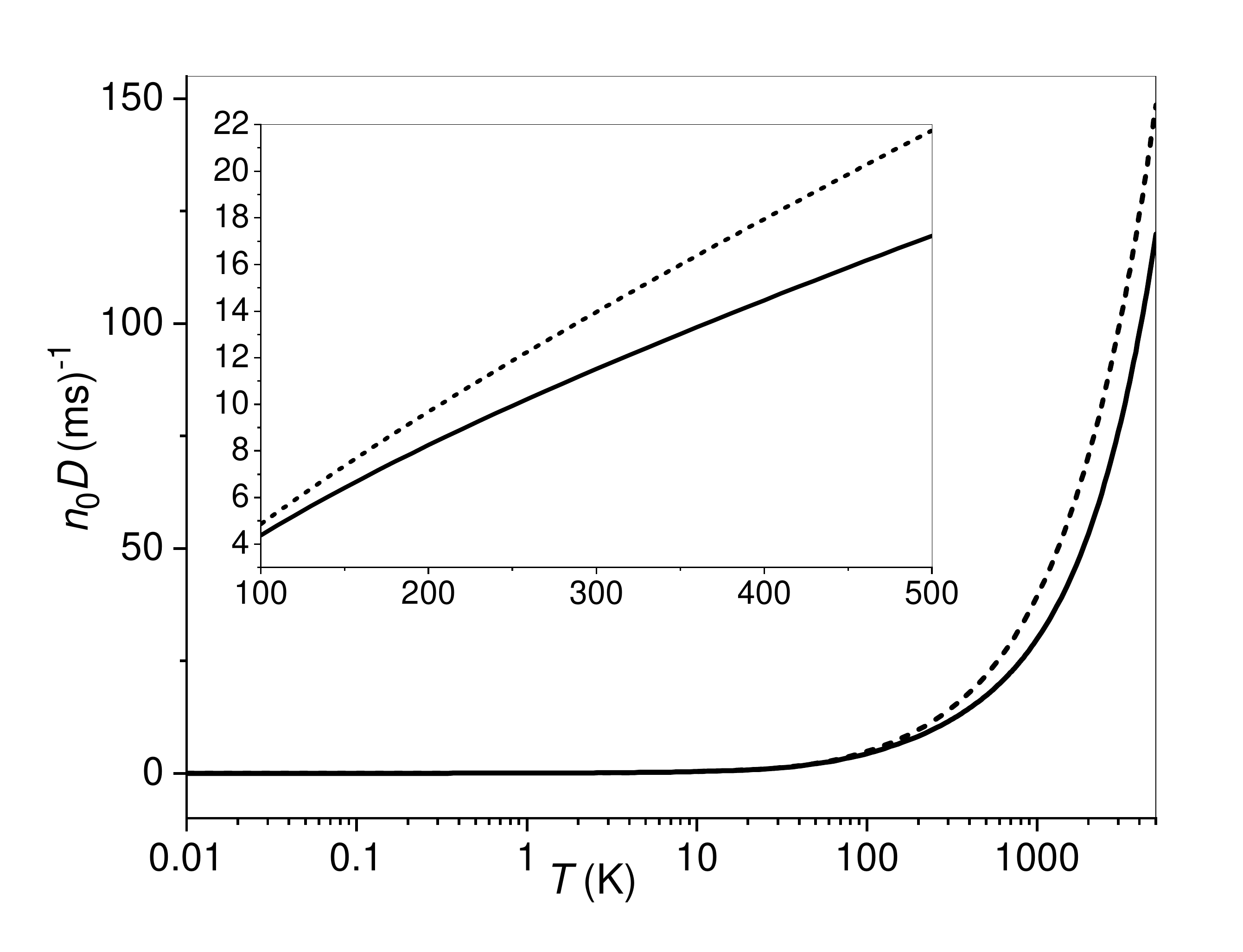}
\caption{\label{fig:DvsT} Zero-field diffusion coefficients $n_0D$
of the Lu$^+$ ions in the ground $X0^+$ state (solid line) and
metastable $^3D_1$ state (dashed line) as functions of
temperature. Inset provides an enlarged view of the $100-500\,$K
region.}
\end{figure}

\begin{figure}
\includegraphics[scale=0.33]{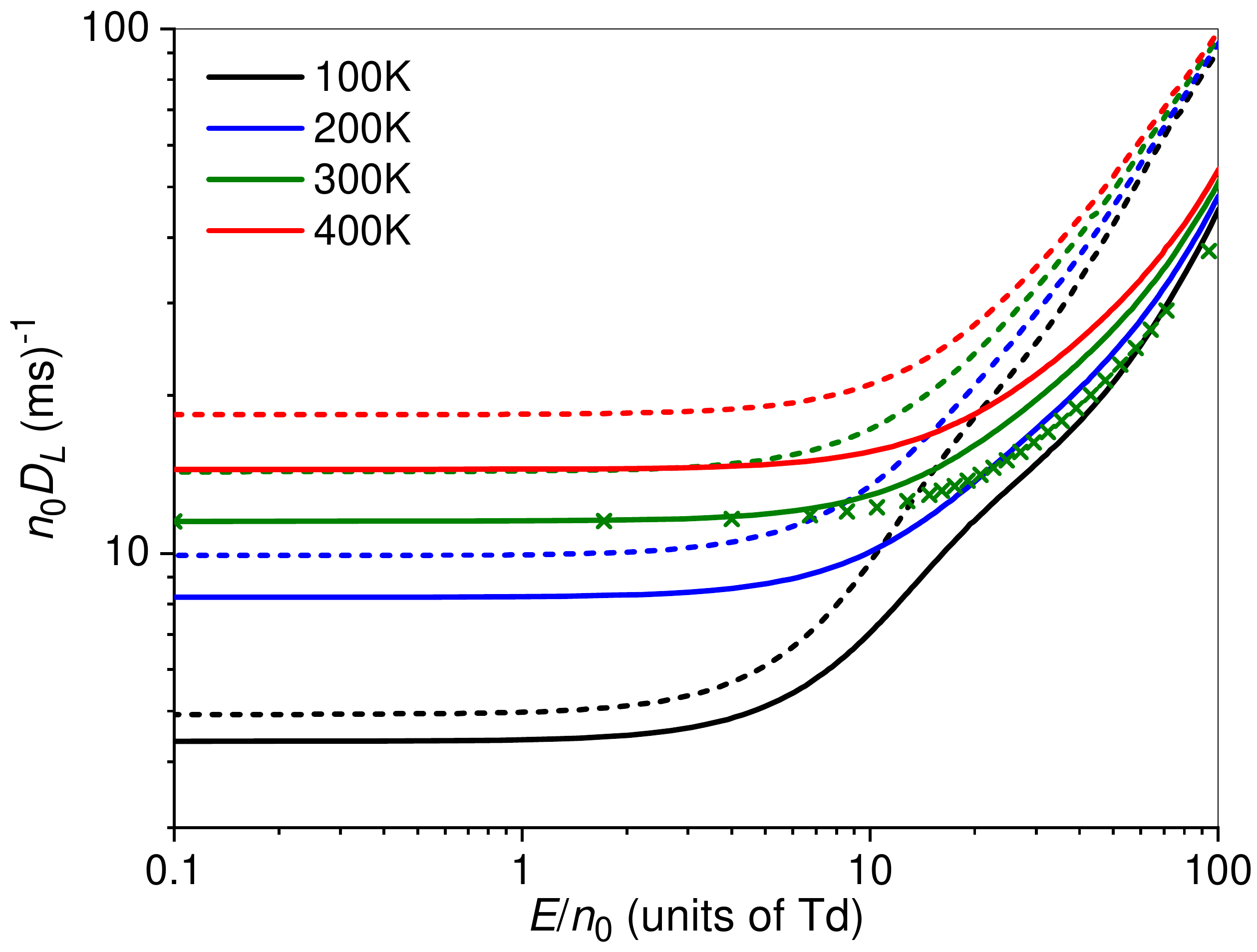}
\caption{\label{fig:Dvsn0} Longitudinal diffusion coefficients
$n_0D_L$ of the Lu$^+$ ions in the ground $X0^+$ state (solid lines)
and metastable $^3D_1$ state (dashed lines) as functions of
$E/n_0$ at selected temperatures. Crosses exemplify the dependence
of the transverse coefficient $n_0D_T$ of the ground-state ion at
$T=300\,$K.}
\end{figure}

We also calculated the diffusion coefficients, as ion diffusion
leads to ion losses and unwanted broadening of drift time distributions.
Zero-field values of the $n_0D$ products are depicted in
Fig.~\ref{fig:DvsT} as functions of temperature. Lower temperatures
minimize both absolute diffusion effects and their difference for
two electronic states. An external electrostatic field distinguishes
longitudinal diffusion along the field direction and transverse
diffusion perpendicular to it. The rapid increase with $E/n_0$ of
the longitudinal diffusion coefficients $D_L$ multiplied by $n_0$ is
shown in Fig.~\ref{fig:Dvsn0}. The transverse coefficients $D_T$
behave similarly. Similar to mobility, strong
variation of the diffusion coefficients takes place at $E/n_0>10\,$Td.

From these transport coefficient calculations we conclude that
metastable-state ions drift faster than the ground-state ones,
but experience more diffusion.

\begin{figure}
\includegraphics[scale=0.33]{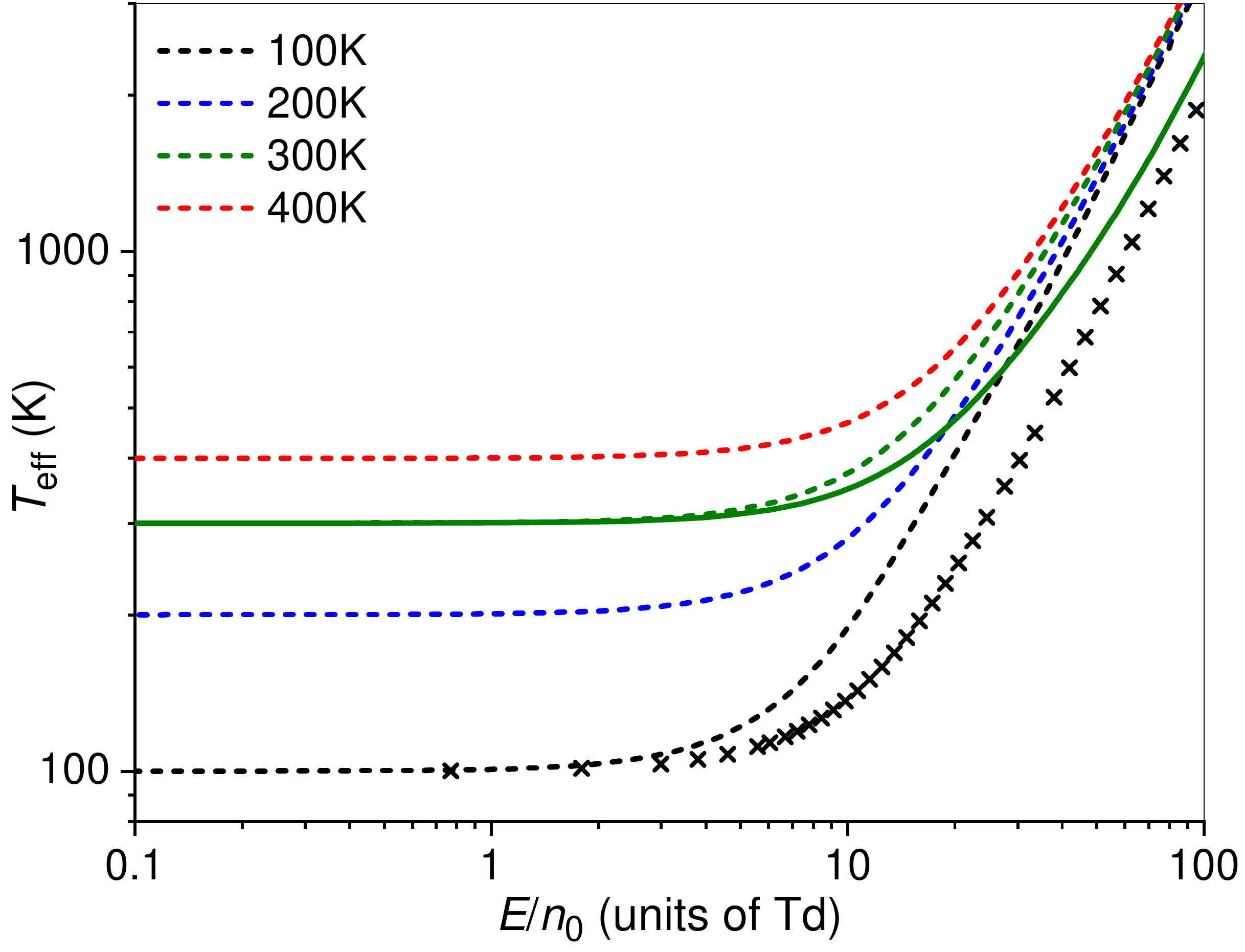}
\caption{\label{fig:effectiveT}
Effective kinetic temperatures $T_\mathrm{eff}$ of
the Lu$^+$($^3D_1$) ions along the drift field at selected
buffer gas temperatures as functions
of $E/n_0$ (dashed lines). Solid line exemplifies the same quantity
for the ground-state Lu$^+$ ion at $300\,$K, while crosses show the
dependence of $T_\mathrm{eff}$ in the direction perpendicular to the
field axis at $100\,$K.}
\end{figure}

Another important parameter is the effective
kinetic temperature of the ion $T_\mathrm{eff}$~\cite{Viehland:2018}.
Being related to the transport coefficients, $T_\mathrm{eff}$ defines
the kinetic energy of the ion-atom collision and hence affects the
rate of collision-induced quenching of excited states. We found
that this stays close to buffer gas temperature $T$ up to
$E/n_0\approx10\,$Td and then rapidly increases; see
Fig.~\ref{fig:effectiveT}.

In order to achieve the best discrimination of ions in the
metastable state a detailed analysis of the drift times is in order.
Bearing in mind the pumping model
introduced before wherein quenching rates for Ba at $880\,$K were
used, we considered $E/n_0$ values up to $40\,$Td in our analysis in
Sec.~\ref{Sec_Signal} to guarantee that $T_\mathrm{eff}$ stays below
$1100\,$K.

\section{Drift time analysis}
\label{Sec_Simulations}
\begin{figure}
\includegraphics[scale=0.33]{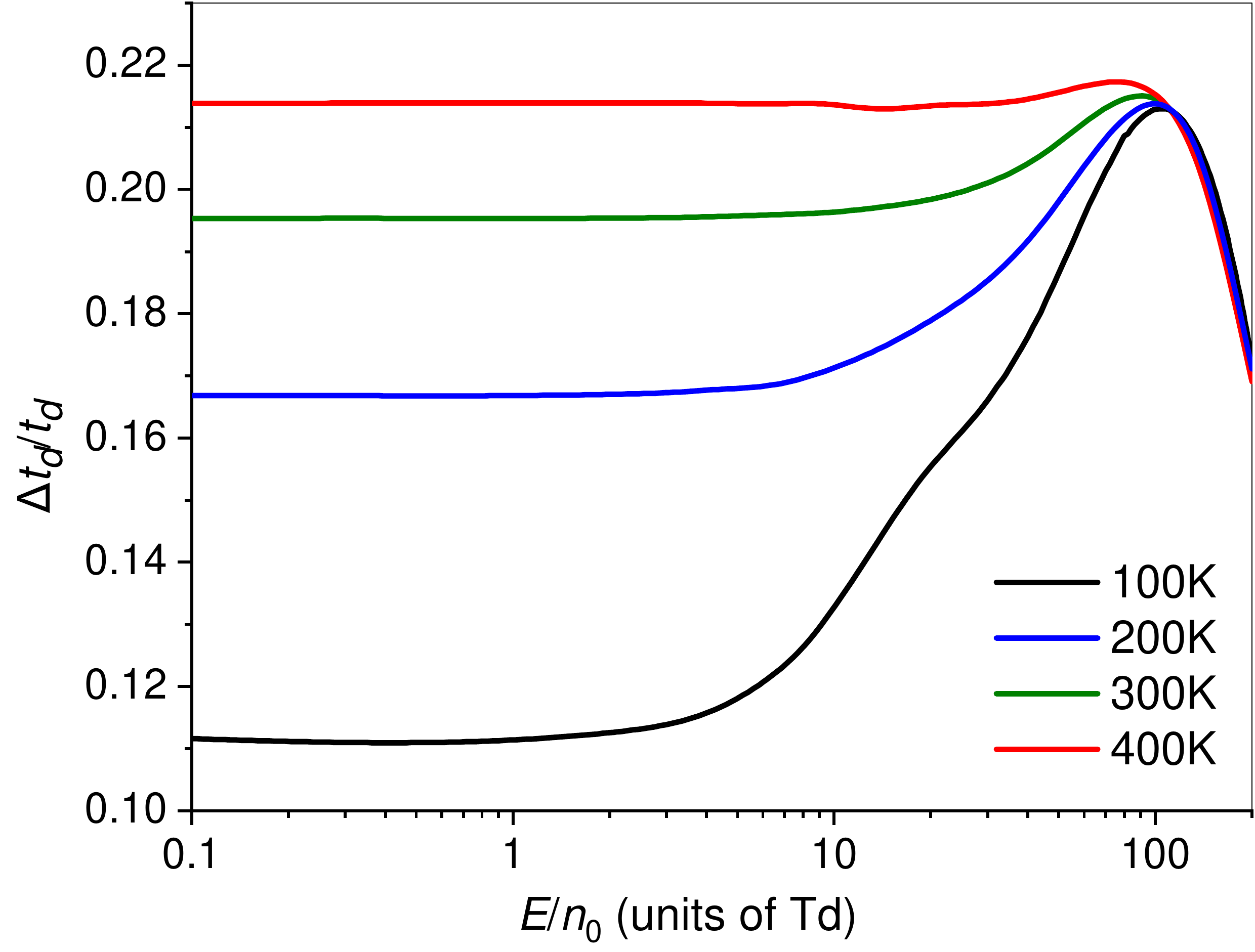}
\caption{\label{fig:Dttd} Relative difference of the drift times for
Lu$^+$ ions in the ground and metastable states.}
\end{figure}

Ion mobility $K$ determines the steady-state drift
velocity of an ion in a buffer-gas filled drift tube under the
influence of a permanent electric field $E$ according to $v_d = KE$.
The mean time needed for an ion swarm to pass through the tube is
\begin{equation}\label{tddef}
t_d=\frac{L}{K\,E}=\frac{L}{N_L\,K_0\,(E/n_0)},
\end{equation}
where the definition of reduced mobility $K_0 = n_0\,K/N_L$ is used
and $N_L$ is the Loschmidt number. The drift time depends on the
drift length $L$, the gas temperature $T$ (through $K_0$ and $n_0$),
the pressure $p_0$ (through $n_0$), and $E/n_0$ (explicitly and
through $K_0$). The difference between the drift times of the
ground-state ions, $t_d$, and of the metastable-state ions, $t^*_d$,
that underlies the electronic state chromatography
effect~\cite{Armentrout:2011} reflects the difference in the
respective mobilities as functions of $T$ and $E/n_0$. It can be expressed as
\begin{equation}\label{deltaT}
\Delta t_d = t_d - t^*_d =
\frac{L}{N_L(E/n_0)}\frac{K^*_0-K_0}{K^*_0K_0}.
\end{equation}
More convenient in the present context is to use the reduced time $t/t_d$, so that
\begin{equation}\label{deltaTtd}
\Delta t_d/t_d=(K^*_0-K_0)/K^*_0
\end{equation}
gives the relative difference of the drift times.
This relative difference can be optimized for best
chromatography performance by finding proper drift conditions of temperature,
pressure, and electric-field strength. The larger the difference the more likely is
detection of resonant pumping from the appearance of a distinct metastable-state
peak in the arrival time distribution. Figure~\ref{fig:Dttd} shows this relative
difference at different $T$ as a function of $E/n_0$. With rising temperature,
$\Delta t_d/t_d$ gradually increases in the low-field region and loses the
dependence on $E/n_0$. The maximum difference of about $21$\%
is achieved either at temperatures above $400\,$K for a wide range of $E/n_0$ or
for $80\lesssim E/n_0\lesssim100\,$Td. Thus, elevated temperatures help
to achieve better discrimination of ions of two types by their mean drift times.
However, maximizing this difference is by far not the only prerequisite for
reliable detection of the laser resonance. Although lowering the temperature down
to $100\,$K and thus reducing the resolution down to a minimum of about $11$\%,
one gains an additional degree of freedom through $\Delta t_d/t_d$ dependence on
$E/n_0$ and a substantial reduction in diffusion and in effective temperature of
the ion.

The mean drift time characterizes the center of the
distribution created by the ion swarm arriving to a detector. The
shape of the distribution depends on the diffusional spread of the
swarm along the field axis, while diffusion in a plane perpendicular
to the field causes ion losses. For more realistic simulations of
these factors we used the analytical expression for the ion flux
derived by Moseley \textit{et~al.}~\cite{Moseley:1969} (see also
Refs.~\cite{Viehland:2018} and \cite{Orient:1974}). The swarm of
ions is assumed to be injected instantaneously from the buncher at
time $t=0$ into the cylindrical drift tube along its axis that
coincides with the external electric-field direction.
This generates an ion flux at a distance $z$ of
\begin{eqnarray}
f(z,t) & = &
\frac{b}{4\pi^{3/2}r_2^2}\frac{z+v_dt}{D_L^{1/2}t^{3/2}}
\exp\left[-\frac{(z-v_dt)^2}{4D_Lt}\right] \nonumber \\ & \times &
\left[1 - \exp\left(-\frac{r_1^2}{4D_Tt}\right)\right], \label{flux}
\end{eqnarray}
where $b$ is the number of ions in the swarm. In what follows, we assume
$b=1$ and use arbitrary units for the ion flux.
In addition to transport
coefficients $D_L$, $D_T$, and $K_0$, Eq.~(\ref{flux}) contains the
entrance and exit orifice radii, $r_1$ and $r_2$. The drift tube
dimensions are assumed to be much larger than these radii such that
Eq.~(\ref{flux}) still holds. This equation disregards initial swarm
spreads in position and velocity, boundary field imperfections and drift
of injected ions before equilibration. However, we expect these minor
effects to be similar for ions in different electronic states and unlikely
to affect the relative ion counts.

\begin{figure}
\includegraphics[scale=0.35]{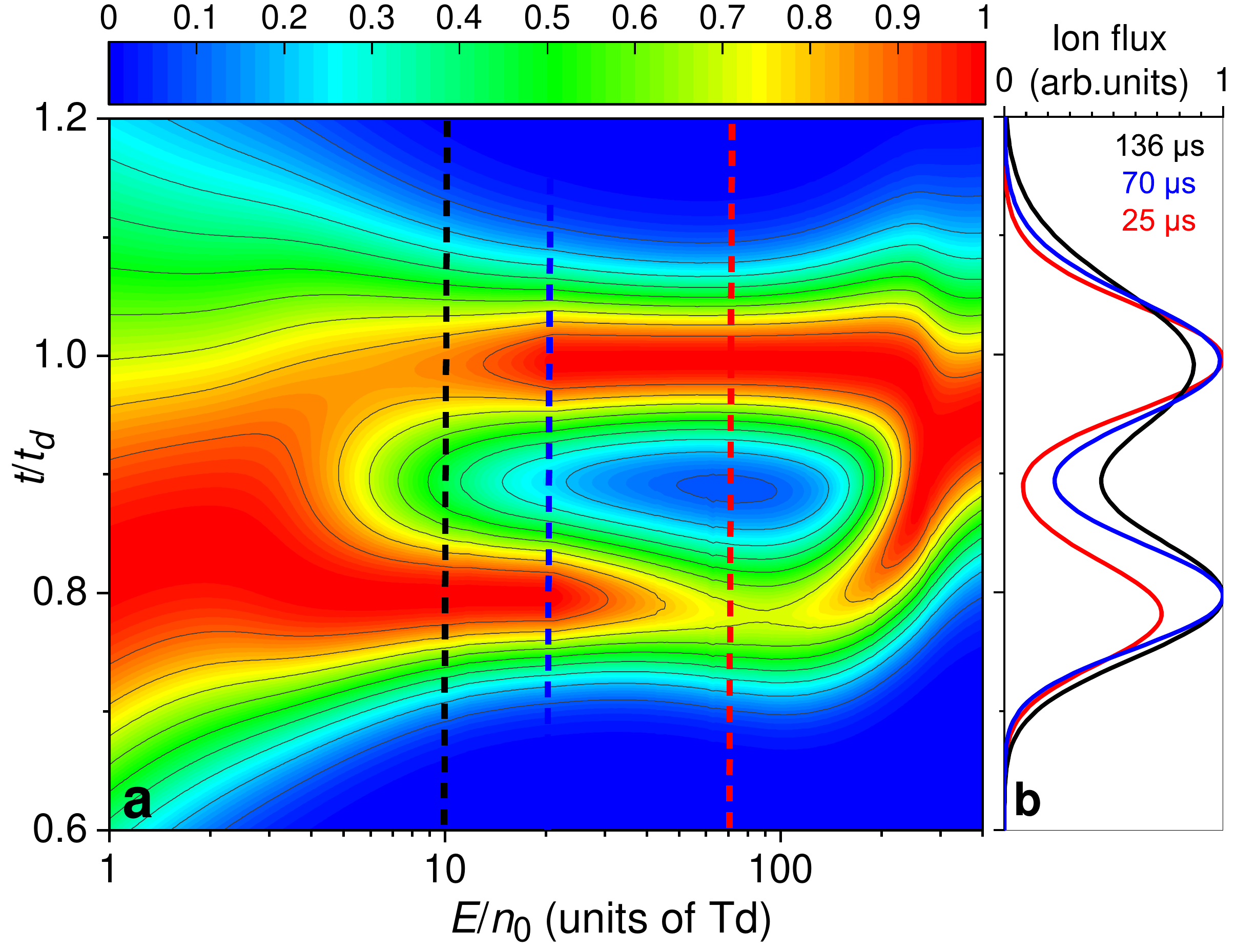}
\caption{\label{fig:map6-1} (a) Relative drift time distribution (color coded) as function of $t/t_d$ and $E/n_0$ for equal initial ensembles of Lu$^+$ in ground and metastable states. Ion losses are neglected.
Horizontal lines mark $E/n_0$ sections depicted in panel (b).
Drift length is $6\,$cm, temperature is $300\,$K, and pressure is
$1\,$mbar. (b) The relative ion fluxes at selected $E/n_0$ values of $10$ (black), $20$ (blue), and $70$\,Td (red).
Legend specifies the corresponding $t_d$ values for the ground-state ions.}
\end{figure}

We first consider the principle way to discriminate
Lu$^+$ ions in the ground and metastable states. As starting
parameters, we chose $r_1=r_2=1\,$mm, $L=6\,$cm and a pressure of
$1\,$mbar following Iceman \textit{et~al.}~\cite{Iceman:2007}, who
successfully discriminated metastable states of Mo$^+$ and W$^+$
ions. For this choice, we calculated the drift time distributions at
the distance $z=L$ and $T=300\,$K for each of the ion states.
Figure~\ref{fig:map6-1}(a) shows the contour plot of the
$f(L,t/t_d)$ distributions assuming a $1:1$ abundance of Lu$^+$ ions
in ground and metastable states depending on $E/n_0$ for the initial parameter choice.
The distributions are normalized to unity at the highest maximum and
thus disregard ion losses. At $5\,$Td
the ground- and metastable-state ions cannot be discriminated and
jointly form a structureless distribution. As ions move faster with
increasing $E/n_0$ values, peak broadening decreases and ions in the
different states can be discriminated, which is the case for $5<E/n_0<200\,$Td.
Figure~\ref{fig:map6-1}(b) shows
$t/t_d$ distributions at selected $E/n_0$ values where two peaks are
clearly visible: one (the ``ground peak,'' $t/t_d=1$) corresponds
to the ground state and another (the ``metastable peak,''
$t/t_d\approx0.8$) corresponds to the metastable-state ions.
A best peak discrimination is achieved at $E/n_0=70\,$Td. However, the
difference in ion mobility becomes smaller with increasing $E/n_0$
values (\emph{cf.} Fig.~\ref{fig:K0vsn0}). Above $100\,$Td the
mean drift times $t_d$ and $t^*_d$ become very close to each other.
In addition, longitudinal diffusion strongly increases in accordance
with Fig.~\ref{fig:Dvsn0} and smooths the distribution even though
mobilities of the ground- and metastable-state ions differ
significantly.

This preliminary consideration indicates that the
ground- and metastable-state ions can be readily discriminated at
room temperature. In view of expected quenching of the metastable
state at elevated $T_\mathrm{eff}$, an optimum discrimination is
desirable at $E/n_0$ values below $40\,$Td. In this case, a reduction
of diffusional spread can be achieved by shortening the drift
length, lowering the temperature (reducing diffusion coefficients),
and/or raising the pressure.

\begin{figure}
\includegraphics[scale=0.4]{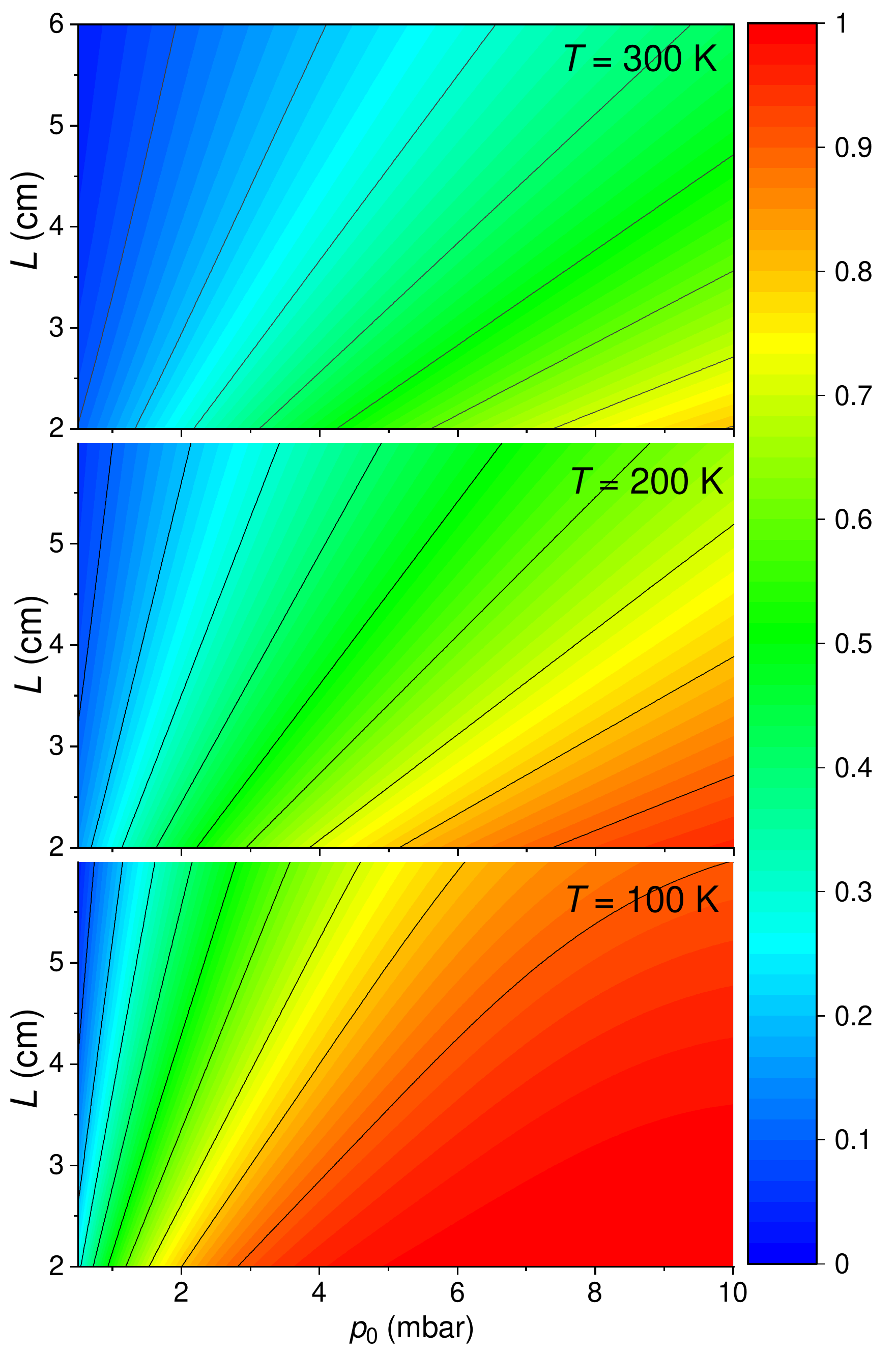}
\caption{\label{fig:transmission} Transmission efficiency of the Lu$^+$($^3D_1$) ions (color coded) as function of $L$ and $p_0$ at
selected temperatures. $E/n_0 = 20\,$Td.}
\end{figure}

For a limited number of ions to be investigated as
proposed in Ref.~\cite{Laatiaoui:2020}, ion losses are the next
major issue. The diffusional spread of the ion swarm in a direction
perpendicular to the electric-field axis strongly affects the efficiency
for transmitting ions through the drift tube. Diffusional spread is
present for ions in both electronic states, being more significant
for the metastable one due to a larger diffusion coefficient; see
Fig.~\ref{fig:Dvsn0}. This is evident from Fig.~\ref{fig:map6-1}(b),
which shows a decrease of the metastable peak intensity with
increasing $E/n_0$. Based on Eq.~(\ref{flux}), we calculated the
transmission efficiency for the metastable Lu$^+$($^3D_1$) ions as a
function of the drift length. As Eq.~(\ref{flux}) does not provide
normalized ion count, we introduced, following Moseley
\textit{et~al.}~\cite{Moseley:1969}, the time-integrated flux
\begin{equation}
A(z)  =  \int_0^\infty f(z,t) dt \label{signal}
\end{equation}
and defined the transmission efficiency as $A(L)/A(z=1\,$mm$)$ at
certain $L$, $E/n_0$ and $p_0$ values. The results are presented in
Fig.~\ref{fig:transmission} as a contour map of efficiency in the
($L$,$p_0$) parameter space at $100$, $200$ and $300\,$K and $E/n_0$
of $20\,$Td. As expected, for a fixed drift path, buffer gas
pressure reduces diffusional losses, but the effect of temperature
is much more profound. Cooling the gas down to $100\,$K allows one
to achieve $90$\% transmission in a wide range of pressures. We also
infer that the drift length of $L=6\,$cm is too long to maintain ion
losses below $10$\% even at elevated
pressures. In general, the shorter the drift length, the smaller
the ion losses. The limit is set by the technical feasibility and
handling and by the initial spatial spread of ions, which could be
magnified due to boundary imperfections of the electric-field
strength and the nonequilibrated ion motion after injection into
the drift tube. A drift length of $L=4\,$cm
appears a reasonable compromise,
allowing one to transmit more than $50$\% of the metastable ions at
$100\,$K and pressures above $1.5\,$mbar. We also stress that
suppressing diffusional losses simultaneously reduces the
diffusional spread, hence improving the discrimination of the ions.

The collection of the metastable ions is also
affected by collisional quenching during the drift. To estimate
the quenching effect conservatively, we repeated the simulations
assuming that the quenched metastable ions are lost. This was
accounted for by multiplying Eq.~(\ref{flux}) by the factor of
$\exp(-\alpha_{51} t)$, where $\alpha_{51}=kn_0$ and $k$ is the
quenching rate constant~\cite{Moseley:1969}. For $k$ we used the
value of 10$^{-13}\,$cm$^3$/s, as determined by Brust and
Gallagher~\cite{Brust:1995} for the metastable Ba($^3D_1$) atoms
in He at $880\,$K; see Sec.~\ref{Sec_Pumping}.
\begin{figure}
\includegraphics[scale=0.4]{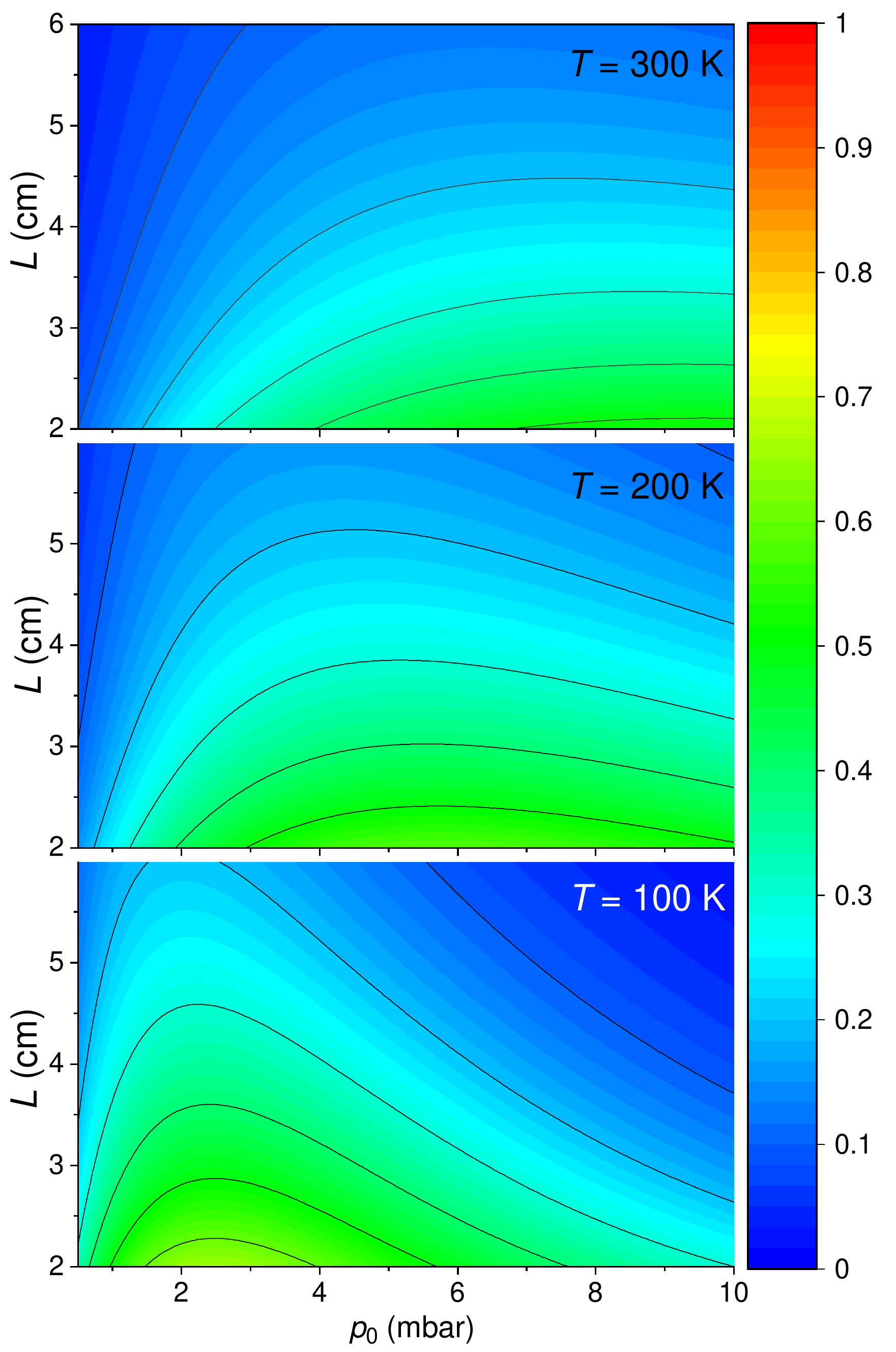}
\caption{\label{fig:signal3} Collection efficiency of the
Lu$^+$($^3D_1$) ions (color coded) as function of $L$ and $p_0$ at
selected temperatures. $E/n_0 = 20\,$Td.}
\end{figure}

The efficiency of metastable ion collection,
defined with the same normalization as the efficiency of ion
transmission, is shown in Fig.~\ref{fig:signal3}. Comparison with
Fig.~\ref{fig:transmission} illustrates the very significant effect
of quenching. At low pressures, the losses are still controlled by
diffusion. As the number of ion-atom collisions increases with
pressure, quenching losses start to dominate. The interplay between
the two effects determines the optimal pressure range. For $T=100\,$K
and $L=4\,$cm, for instance, $35$\% of the metastable ions can be
collected at the exit orifice of the drift tube at pressures between
$1.5$ and $4\,$mbar.

\section{Resonance detection by ion count}\label{Sec_Signal}
In the simulations presented above we identify the range
of experimental conditions at which the ions in the metastable
states can be discriminated with minimum losses. Accurate simulation
of the ion count signal to be detected would require explicit consideration
of collisional coupling between the electronic states of ions during
the drift. It goes beyond the present analytical picture as the
equilibrium between two states is achieved at much longer times than
the drift itself. Iinuma and co-workers considered
the drift of ions coupled by fast reversible reactions (see
Ref.~\cite{Iinuma:1994} and references therein) but this approach is
inappropriate for slow irreversible quenching and reduces to ground-state
ions only. We therefore assume the quenching to proceed too slowly
to affect the drift time of ions in both states and can only alter
their amount by increasing the number of the ground-state ions to
the same extent as it decreases the number of the metastable ions.
The flux of the ions in the metastable state is therefore
represented by Eq.~(\ref{flux}) multiplied by $C_m\exp(-\alpha_{51}
t)$, where $C_m$ is the initial fraction of the metastable ions in
the swarm. The same equation multiplied by $[1-\exp(-\alpha_{51}
t)]C_m + (1-C_m)$ represents the flux of the ground-state ions.
\begin{figure}
\includegraphics[scale=0.35]{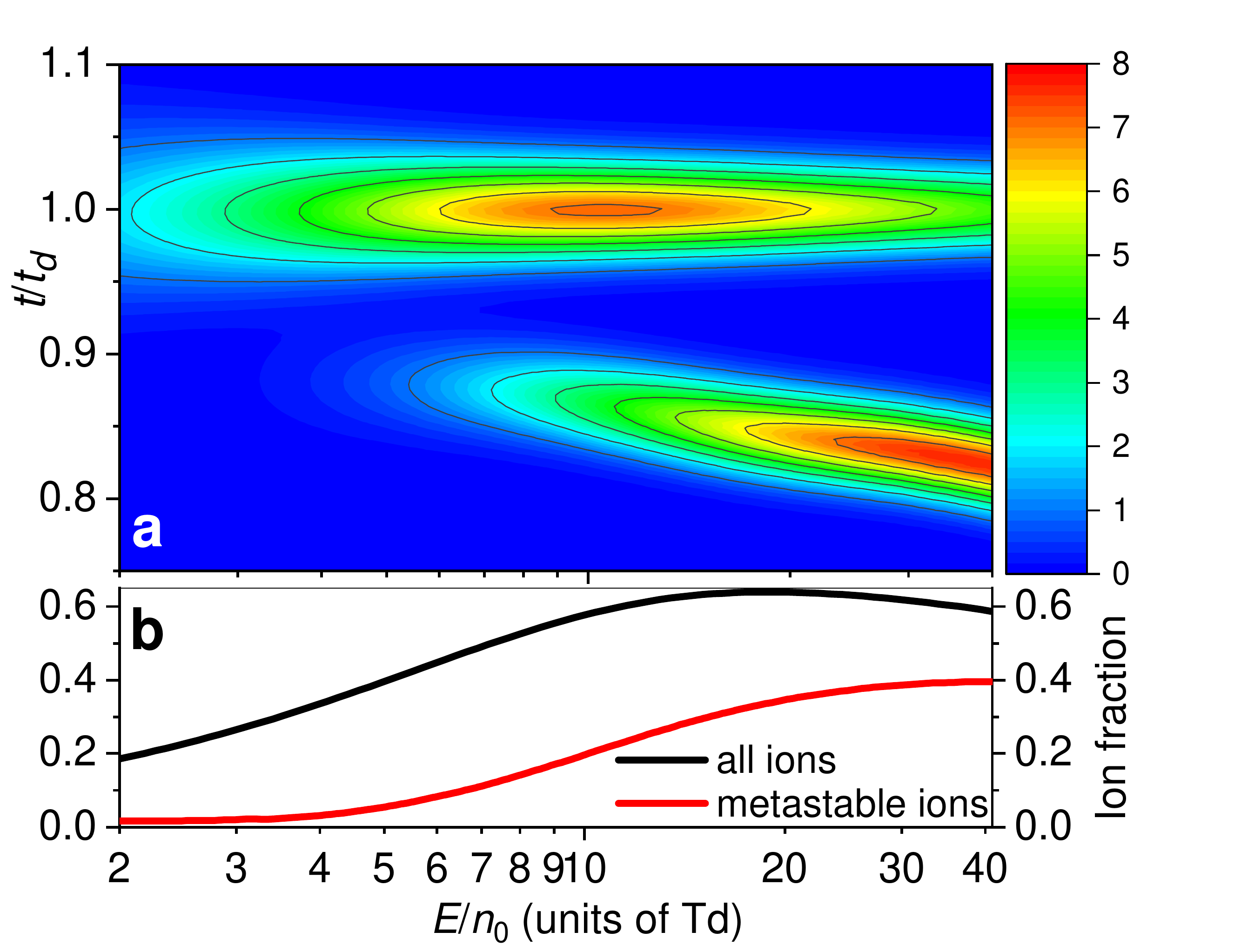}
\caption{\label{fig:tdresmap} (a) Ion signal at resonant pumping
$I_p(t)$ (color coded, arbitrary scale) as function of
$t/t_d$ and $E/n_0$. (b) The
fractions of collected ions, all and metastable. For both panels,
$T=100\,$K and $p_0=2\,$mbar.}
\end{figure}

With these weighted functions it is now possible to
model expected ion signals with off-resonance and resonant pumping
proceeding via Lu$^+$ $^1S_0$--$^3P_1$ excitations. We define
the former as $I_g(t)=f_g(L,t)$, where $f_g$ is calculated for the
swarm of the ground-state ions and $L=4\,$cm. The signal at resonant pumping
$I_p(t)=f_p(L,t)$ is represented through the combined flux
function that accounts for initial ground and metastable state
populations of the ions released from the buncher, as well as the
population loss and gain due to quenching in the course of the
drift. As follows from the population kinetics in the buncher with
the residual He pressure of $5\times10^{-2}\,$mbar (see
Sec.~\ref{Sec_Pumping}), $94$\% of the ground-state population in Lu$^+$
is transferred to the $^3D_1$ metastable state at resonance after
ten laser pulse exposures. Accordingly, we simulated the signals
$I_g(t)$ and $I_p(t)$ by taking $C_m=0.94$. A small $^1D_2$
occupation fraction predicted by the kinetic model was added to the
ground-state population even though its feeding of the lower-lying
$^3D_J$ metastable states is more likely due to efficient
intramultiplet collisional relaxation~\cite{Brust:1995}.
Quenching of metastable states due to ion-ion interactions
has been neglected in view of small ion quantities to be expected in a single bunch.
Both signals are normalized to the initial integrated ion flux $A_0$
taken at $z=1\,$mm according to Eq.(\ref{signal}) for the
off-resonance flux and thus represent the corresponding fractional
ion counts.

Figure~\ref{fig:tdresmap}(a) shows the signal for resonant pumping
$I_p(t)$ at $T=100\,$K and $p_0=2\,$mbar as a function of $E/n_0$.
Though the signal intensities in the
$t/t_d$-$E/n_0$ map are not
representative, the signal evolution is transparent. At low $E/n_0$
almost all the metastable ions are lost due to the transverse
diffusion and quenching. Despite their minority, the ground-state
ions with $t/t_d=1$ are still transmitted, having smaller diffusion
coefficient and gaining population from quenching. The metastable
ions appear at $E/n_0\approx5\,$Td, when dragging force overpowers
diffusional spread and makes the quenching probability smaller.
Their signal is very well separated in time and grows with $E/n_0$.
At about $20\,$Td, diffusion losses start to manifest themselves
again, as the diffusion coefficients rapidly increase; see
Fig.~\ref{fig:Dvsn0}. The fraction of collected metastable ions is
determined by integrating the $I_p(t)$ signal from $t=0$ to
$(t_d+t_d^*)/2$ and is shown in Fig.~\ref{fig:tdresmap}(b) together
with the fraction of all ions detected.

\begin{figure}
\hfill\includegraphics[scale=0.33]{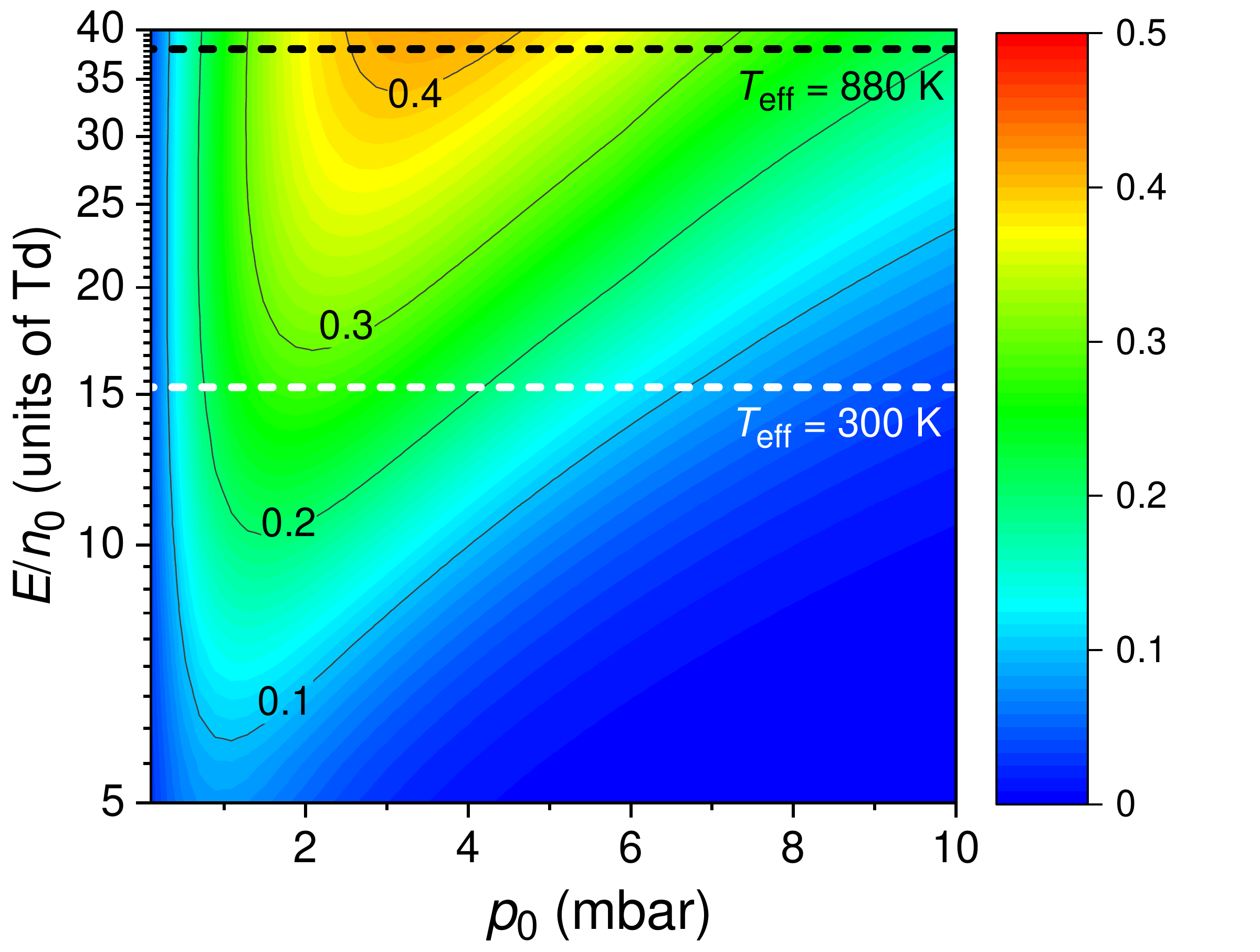}
\caption{\label{fig:metfrac} Fraction of the metastable
ions collected after the drift as function of $p_0$ and $E/n_0$ at $T=100\,$K. Horizontal lines indicate $E/n_0$ values, which correspond to effective temperatures of the metastable ions of $300$ and $880\,$K.}
\end{figure}

Dependence of the fraction of the metastable ions on
both $p_0$ and $E/n_0$ is mapped in Fig.~\ref{fig:metfrac}. Increasing
the latter and tuning the pressure to optimum, one can reach a collection
efficiency far above $40$\%. However, increased ion heating from the
application of elevated $E/n_0$ ratios may promote quenching to the ground
state. By constraining the effective ion temperatures below $880\,$K, we
predict a highest collection efficiency of about $41$\% at $E/n_0=38\,$Td
and $p_0=3.5\,$mbar. A safer presumption of $300\,$K effective
temperature results in a collection efficiency of $28$\%
($E/n_0\approx15\,$Td, $p_0\approx2\,$mbar).
\begin{figure}
\includegraphics[scale=0.33]{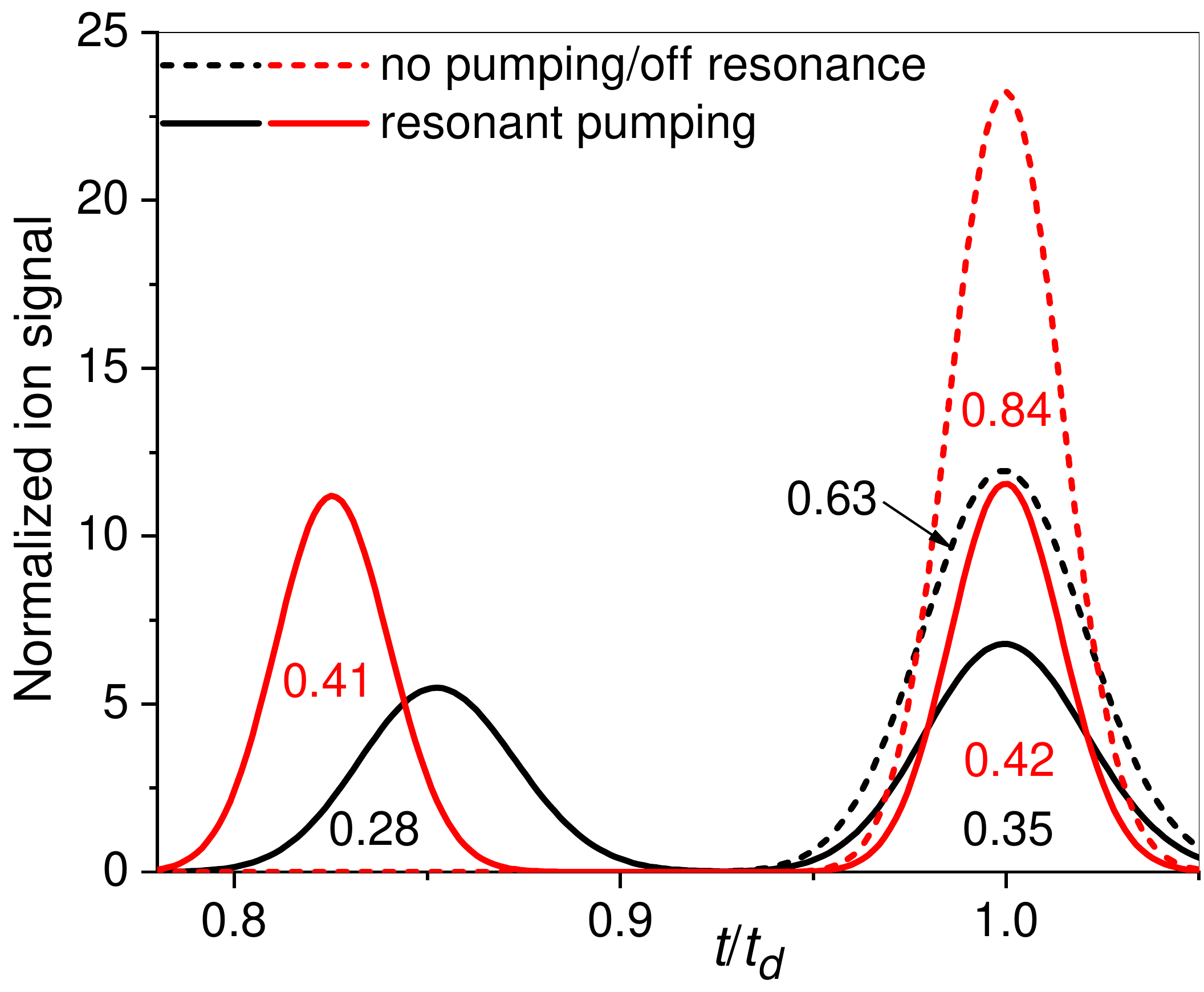}
\caption{\label{fig:findist} Off resonance and resonant
pumping signals $I_g$ and $I_p$, respectively, at optimum drift conditions and effective temperatures of metastable ions of $880\,$K
(red, $p_0=3.5\,$mbar, $E/n_0=38\,$Td) and $300\,$K
(black, $p_0=2\,$mbar, $E/n_0=15\,$Td). $L=4\,$cm and the He temperature is $T=100\,$K. Indicated numbers are the respective fractions of ions collected after the drift (areas of the peaks multiplied by $t_d$).}
\end{figure}

The expected signals at these conditions are presented
in Fig.~\ref{fig:findist}. This shows a complete separation of the peaks for
the ground and metastable ions, as well as a reasonably high collection
efficiency. It should be noted that a more practical way to detect resonant
pumping is to use the difference signal
\begin{equation}
\Delta I(t)=\frac{1}{2}[I_p(t)-I_g(t)]/A_0, \label{ResonanceSignal}
\end{equation}
corresponding to a resonance detection efficiency, where the factor $1/2$
accounts for the fact that the difference signal is formally taken from two ion bunches.
For the drift time distribution shown in Fig.\ref{fig:findist}, it does not
provide any gain in sensitivity, in general.
But, given stable experimental conditions while searching for the resonance,
the reference signal is collected numerous times such that the difference
signal becomes more sensitive to changes in the arrival time distributions,
i.e., the efficiency doubles by neglecting $1/2$ in Eq.~(\ref{ResonanceSignal}).

As discussed in Sec.~\ref{Sec_Simulations}, quenching to the ground state significantly impacts the collection efficiency of metastable ions.
Since the quenching is related to crossings of potential-energy curves~\cite{Armentrout:2011,Brust:1995},
%Since the quenching rate is in general inversely proportional to the energy gap, as discussed for fine-structure levels in Ref.~\cite{Krause:1972},
we expect this rate to be smaller and the metastable ion collection to be larger for Lu$^+$ drifting in He due to the larger gap in Lu$^+$ ($11796\,$cm$^{-1}$) compared with Ba ($9034\,$cm$^{-1}$).

For a Lr$^+$-He system, we expect quenching due to coupling of the potential-energy curves to be much more suppressed, in particular, at low gas temperatures and $E/n_0$ ratios because (unlike in Lu$^+$ and neutral Ba) the $^3D_1$ state in Lr$^+$ is predicted to be far above the ground state ($20846\,$cm$^{-1}$)~\cite{Kahl:2019}.

Nevertheless, if we assume that the case of lawrencium differs
only in pumping kinetics, the resonant pumping signal from the
metastable Lr$^+$ ions would be weaker by the initial population
ratio of $C_m($Lr$^+)/C_m($Lu$^+)=0.56$; see Sec.~\ref{Sec_Pumping}.
A collection efficiency between $16$\% and $23$\% can then be expected
at optimal conditions, which is sufficiently high for the envisaged studies.

Although other ionic species may exhibit ground-state fine structures compared with Lu$^+$ and Lr$^+$, we expect optical pumping to remain efficient in these systems due to intramultiplet collisional relaxation. Similar considerations apply for dark hyperfine components of ground states, which we expect to not play a significant role as power broadening of spectral lines is often beneficial and intended during level searches utilizing broadband lasers.

\section{Summary and Conclusion}
We have modeled optical pumping in Lu$^+$ and
Lr$^+$ ions and calculated the interaction potentials for Lu$^+$ in
its ground ($^1S_0$) and excited ($^3D_1$) state in He. We
predict the mobility to be distinct for the two ionic states at
temperatures $\gtrsim100\,$K, which can be exploited for electronic
state chromatography. These calculations and the description of ion
drift have enabled us to elucidate parameter spaces of
($T$,$E/n_0$), ($L$,$p_0$), and ($E/n_0$,$p_0$) for the best
chromatography performance. We found that optical pumping from
the ground state would lead to a relatively high collection
efficiency of ions in the $^3D_1$ state at optimal conditions. We
expect a state-specific control of the ion transport to be feasible
for both ions, Lu$^+$ and Lr$^+$. Its usage in conjunction with
resonant laser excitations, can be exploited to study the electronic
structure of Lr$^+$ for which so far only theoretical predictions
exist.

\begin{acknowledgments}
This project has received funding from the European Research Council
(ERC) under the European Union's Horizon 2020 research and
innovation programme (Grant Agreement No. 819957). A.A.B. acknowledges the
support from the Russian Foundation for Basic Research (Project No.
19-03-00144).
\end{acknowledgments}

\end{document}